\documentclass[twoside,a4paper,11pt]{sea9}
\usepackage{graphicx}
\usepackage{hyperref}
\usepackage{movie15}
\newcommand{\Funits}{10$^{-16}$ erg~s$^{-1}$~cm$^{-2}$}

\begin{document}
\pagenumbering{arabic}
\pagestyle{myheadings}
\thispagestyle{empty}
{\flushleft\includegraphics[width=\textwidth,bb=58 650 590 680]{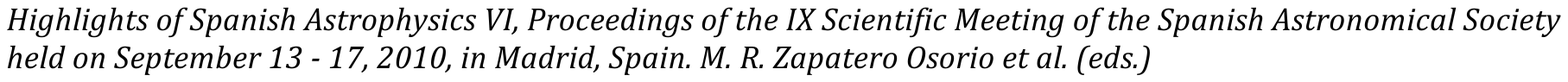}}
\vspace*{0.2cm}
\begin{flushleft}
{\bf {\LARGE
%
CALIFA, the Calar Alto Legacy Integral Field Area survey: Early Report.
%
}\\
\vspace*{1cm}
%
S.F.S\'anchez$^{1}$, 
R.C. Kennicutt$^{2}$,
A. Gil de Paz$^{3}$,
G. Van den Ven$^{4}$,
J.M. Vilchez$^{5}$,
L. Wisotzki$^{6}$,
J. Walcher$^{6}$,
R.A. Marino$^{1,3}$,
E. M\'armol-Queralt\'o$^{1,3}$,
D. Mast$^{5,1}$,
K. Viironen$^{1}$
and
The CALIFA Team$^{7}$
%
}\\
\vspace*{0.5cm}
%
$^{1}$
 Centro Astron\'omico Hispano Alem\'an, Calar Alto, (CSIC-MPG),
   C/Jes\'us Durb\'an Rem\'on 2-2, E-04004 Almeria, Spain\\
$^{2}$
 Institute of Astronomy, University of Cambridge, Madingley Road, Cambridge, UK \\
$^{3}$
 Departamento de Astrof\'{i}sica y CC. de la Atm\'{o}sfera, Universidad Complutense de Madrid, Madrid 28040, Spain \\
$^{4}$
 Max Planck Institue for Astronomy, Königstuhl 17, D-69117 Heidelberg, Germany \\
$^{5}$
Instituto de Astrof\'\i sica de Andaluc\'\i a (CSIC), Camino Bajo de Huetor
 s/n, Granada \\
$^{6}$
 Astrophysical Institute Potsdam, An der Sternwarte 16, D-14482, Postdam, germany\\
$^{7}$ http://www.caha.es/CALIFA/
%
\end{flushleft}
%
\markboth{
CALIFA survey
}{ 
%
S. F. S\'anchez et al.
%
}
\thispagestyle{empty}
\vspace*{0.4cm}
\begin{minipage}[l]{0.09\textwidth}
\ 
\end{minipage}
\begin{minipage}[r]{0.9\textwidth}
\vspace{1cm}
\section*{Abstract}{\small
%
We present the Calar Alto Legacy Integral Field Area survey
(CALIFA). CALIFA's main aim is to obtain spatially resolved
spectroscopic information for $\sim$600 galaxies of all Hubble types in 
the Local Universe (0.005$<z<$0.03). 
The survey has been designed to allow three key measurements to be made: 
(a) Two-dimensional maps of stellar populations (star formation histories, chemical 
elements); (b) The distribution of the excitation mechanism and element abundances 
of the ionized gas; and (c) Kinematic properties (velocity fields, velocity dispersion), 
both from emission and from absorption lines. 
To cover the full optical extension of the target galaxies (i.e. out to a 3$\sigma$ 
depth of $\mu\sim$23 mag/arcsec$^2$), CALIFA uses the exceptionally large 
field of view of the PPAK/PMAS IFU at the 3.5m telescope of the Calar Alto 
observatory. We use two grating setups, one covering the wavelength 
range between 3700 and 5000 {\AA} at a spectral resolution $R\sim$1650, and the other 
covering 4300 to 7000 {\AA} at R$\sim$850. 
The survey was allocated 210 dark nights, distributed in 6 semesters and
starting in July 2010 and is carried out by the CALIFA collaboration, 
comprising $\sim$70 astronomers from 8 different countries. 
As a legacy survey, the fully reduced data will be
made publically available, once their quality has been verified. We showcase 
here early results obtained from the data taken so far (21 galaxies).

%
\normalsize}
\end{minipage}
%
%
%
\section{Introduction \label{intro}}

Much of our recently acquired understanding of the architecture of the
Universe and its constituents derives from large surveys 
(e.g., 2dFGRS, SDSS, GEMS, VVDS, COSMOS to name but a few). Such surveys
have not only constrained the evolution of global quantities such as the
cosmic star formation rate, but also enabled us to link this with the
properties of individual galaxies -- morphological types, stellar masses,
metallicities, etc.. Compared to previous possibilities, the major advantages of
this recent generation of surveys are: (1) the large number of objects
sampled, allowing for meaningful statistical analysis to be performed on an
unprecedented scale; (2) the possibility to construct large comparison/control
samples for each subset of galaxies; (3) a broad coverage of galaxy subtypes
and environmental conditions, allowing for the derivation of universal
conclusions; and (4) the homogeneity of the data acquisition, reduction and
(in some cases) analysis.

\begin{figure}[t]
\center
\includegraphics[width=15.5cm,trim=20 252 56 228,clip=true]{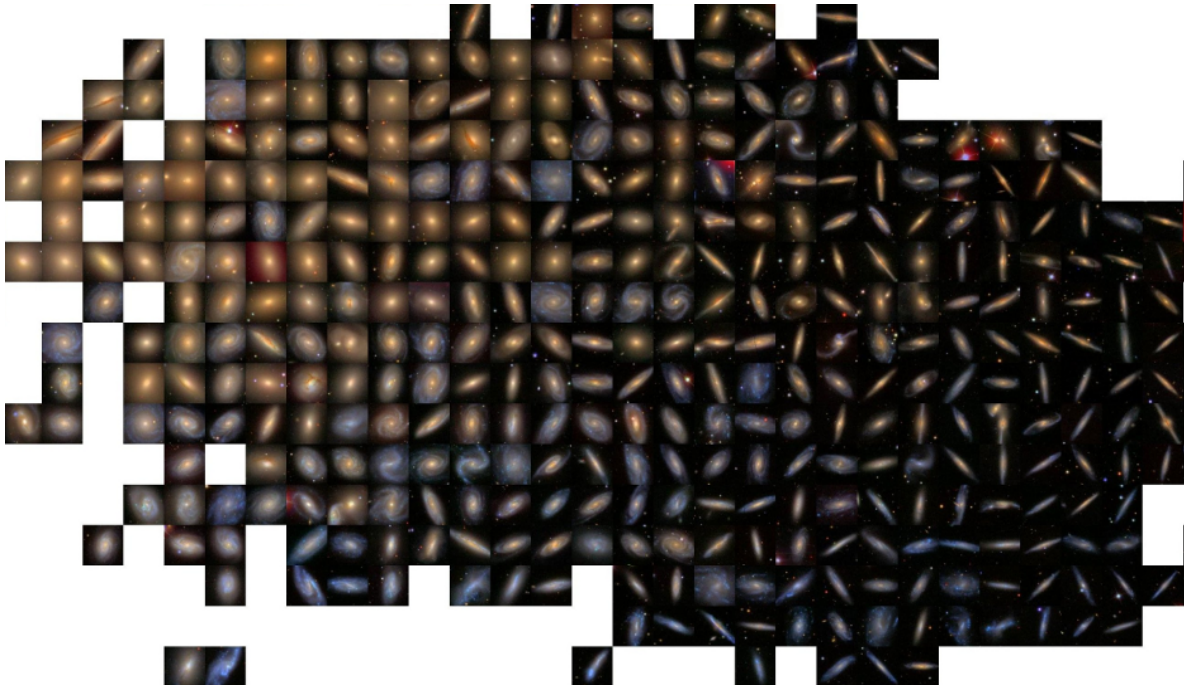} ~
\caption{\label{fig1} Postage stamp (90''$\times$90'') true-color images of a subset 
of galaxies within the CALIFA mother sample , extracted from the SDSS
  dataset, ordered following the $u-r$ vs. $r$ color-magnitude diagram. The
  figure spans from M$_r\sim$-23 mag from the left end, to M$_r\sim$-18 mag
  to the right end, and from $u-r\sim$3.5 mag from the top end, to $u-r\sim$1.5
  mag to bottom end. The figure illustrates the large variety of galaxy types
  covered by the survey.
}
\end{figure}

An observational technique combining the advantages of imaging and
spectroscopy (albeit with usually quite small field of view) is Integral Field
Spectroscopy (IFS). However, so far this technique has rarely been used in a
`survey mode' to investigate large samples, with a few notable exceptions
(e.g., SAURON, de Zeeuw et al. 2002). On the other hand, the large single 
fiber surveys mentioned above have limitations that can only be overcome 
by a statistical sample of nearby galaxies with spatially resolved spectroscopic 
information. In order to address this requirement, we proposed the
CALIFA survey. This survey has been granted with 210 dark nights of the 3.5m
telescope at Calar Alto Observatory (Spain), homogeneously distributed along 6
semesters, officially starting the 1st of July 2010.  CALIFA will observe a
well-defined sample of $\sim$600 galaxies in the local universe with the
PMAS/PPAK integral field spectrophotometer (Roth et al. 2005; Kelz et
al. 2006), mounted on the 3.5 m telescope at the Calar Alto Observatory
(Spain). The sample to be observed was selected to comprise most galaxy
types, covering the full color-magnitude diagram down to M$_B<-$18 mags.  The
observations will cover the optical wavelength range between 3700 and 7000\AA,
using two overlapping setups, with resolutions of R$\sim$1650 and
R$\sim$850. Considering this spectral coverage, and the large field-of-view of
PPAK ($>$1 arcmin$^2$), CALIFA is thus the largest and the most comprehensive
wide-field IFU survey of galaxies carried out to date.

\begin{figure}
\begin{center}
\includegraphics[width=12cm]{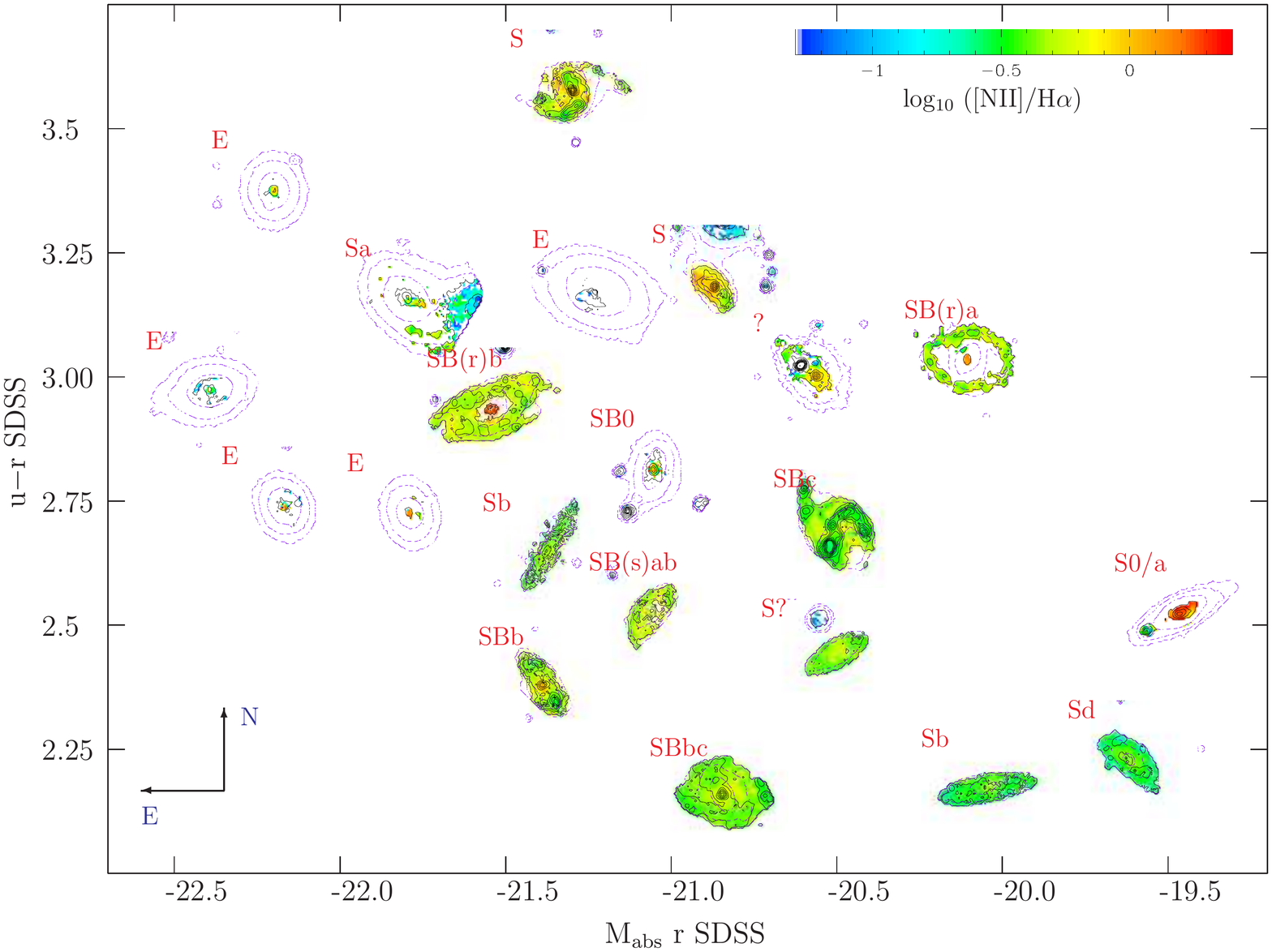}
\caption{\label{fig:fig_gas_NII} Similar color-magnitude diagram as presented
  in Figure 1, with the color-maps showing the distribution of the
  emission line ratios between [NII]$\lambda$6583 and H$\alpha$, derived by
  the fitting procedure. In this case the solid-contours show the intensity of
  the H$\alpha$ emission. The dashed-blue-contours show 3 intensity levels of
  the continuum emission at $\sim$6550\AA\  (starting at 3\Funits). They have
  been included to indicate the physical extension of the continuum emission
  in the galaxies.}
\end{center}
\end{figure}

\section{Science Goals}

One of the most fundamental challenges in astrophysics is to understand the
origin for the observed diversity of galaxies, and the physical mechanisms --
intrinsic and environmental -- that are responsible for the differences as
well as similarities between them. Detailed studies of nearby galaxies can
help by revealing structural properties that can be interpreted as ``fossil
records'' of the formation and evolution process.  We have long known from our
own Milky Way that there are intricate links between chemical and kinematic
properties of stellar populations, as well as between stars and gas,
and similar relations have been found in other galaxies. An old but still
unanswered question is the problem of ``nature vs. nurture'', i.e. the
relative importance of environmental processes such as merging and accretion,
relative to intrinsic secular processes that inevitably occur in an evolving
complex dynamical system. A more recently posed puzzle is the origin of 
the bimodality of the galaxy population. What is it that makes galaxies be either 
``red and dead'' or "blue and star forming", and in particular, what is happening 
to galaxies in the intermediate ``green valley'' of the color-magnitude diagram? 
An important contribution to progress in these areas can come from 
spatially resolved spectroscopy of a statistical sample of galaxies, such as will be 
provided by the CALIFA Survey. In the following we list a few specific questions 
that CALIFA will allow us to address: 

\begin{itemize}

\item Characterization of galaxies over their \emph{full} spatial extent, i.e. avoiding 
aperture biases and harnessing the additional power of 2D resolution (gradients, 
anomalies). 

\item The environmental dependence of the stellar populations in the \emph{outskirts} 
of galaxies, where they are most likely to be affected (e.g.~origin of disk truncation). 

\item Gas ionization mechanisms (star formation, shocks, AGN) in dependence on 
location inside the galaxy and galaxy overall properties. 

\item Kinematic classification of galaxies of \emph{all} Hubble types, e.g.~are there 
dichotomies similar to the slow vs. fast rotators also in late type galaxies?

\item Chemical evolution of \emph{entire} galaxies and the origin of the scaling 
laws of gas metallicity (through e.g.~abundance gradient dependance on internal 
and external properties). 

\item The nature of the galaxies in the green valley, e.g.~ whether galaxies stop forming 
stars from the outskirts or from inside out. 

\item Finally, comparison to full chemo-dynamical models, i.e. weighing in the relative 
importance of dynamical processes (bars, minor mergers, migration etc.) and star formation 
processes (feedback, stellar evolution, etc.) for the chemical enrichment processes in galaxies.

\end{itemize}

To address these scientific questions, we have designed the survey so it allows 
us to make three key measurements: 
(a) Two-dimensional maps of stellar populations (star formation histories, chemical 
elements); (b) The distribution of the excitation mechanism and element abundances 
of the ionized gas; and (c) Kinematic properties (velocity fields, velocity dispersion), 
both from emission and from absorption lines. 
All these quantities will be reconstructed in maps covering the entire
luminous extent of the galaxies in the sample, a first in galaxy evolution studies.

\section{Sample}

The CALIFA mother sample has been tailored to fulfill the main requirements of
the science goals of the survey, i.e., the characterization of the spatially
resolved spectroscopic properties of galaxies in the Local Universe of any
kind, on one hand, and to maximize spatial coverage of the IFU over the
complete size of the galaxies, on the other. Based on these requirements and 
to guarantee good photometric coverage, the
mother sample has been selected from the SDSS DR7 photometric catalogue
(Abazajian et al. {2009}), adopting a combination of angular isophotal diameter 
selection ($45'' < D25 < 80''$) with redshift selection ($0.005<z<0.03$). 
The final mother sample comprises $\sim$1000 galaxies, and is a 
representative subsample of the galaxies in the Local Universe (Mast
et al., in prep). The final observed sample by CALIFA will be selected from
this mother sample, based on the visibility for each night. It is expected
that this will produce a random subsample of $\sim$600 galaxies.

Figure 1 shows postage stamp images of a random selection of galaxies within the
mother sample, distributed along the color-magnitude diagram, illustrating the
diversity of galaxies and the wide range of parameters covered by the survey
($\sim$5 magnitudes in luminosity and $\sim$3 in color).

\section{State of the Survey}

CALIFA has recently started. Its data acquisition phase will last for three
years, and we have just got data for 21 galaxies (June-July 2010). The reduction
of the CALIFA data is performed using a fully automatic pipeline, that 
operates without human intervention, producing both the scientificaly useful
frames and a set of quality control measurements that help to estimate the
accuracy of the reduced data. The pipeline uses the routines included in the
{\sc R3D} package (S\'anchez et al. 2006) and the {\sc E3D} visualization tool
(S\'anchez et al. 2004). The reduction consists of the standard steps for
fibre-based integral-field spectroscopy.

The first acquired data have been fully reduced using the implemented
pipeline. To determine their quality, a set of exploratory analyses
were performed in order to derive the three key measurements that drive 
the survey: stellar population properties, ionized gas properties, and 
kinematic information in both components. For this exploratory analysis we 
adopted procedures based on FIT3D (S\'anchez et al. 2007), described in 
detail in S\'anchez et al. (2010), and used in previous surveys (eg., PINGS,
Rosales-Ortega et al. 2010). It is expected that the final analysis of the CALIFA 
data will also use a large number of other tools, both from within the collaboration 
and from the scientific community. 

Figure 2 showcases one of the outputs of this analysis. We show there 
the spatial distribution of the [NII]/H$\alpha$ emission line ratio (a
classical ionization diagnostic parameter), for the different observed galaxies,
distributed along the the color-magnitude diagram. This figure illustrates the
kind of comparative analysis that can be performed with CALIFA, where different
spatially resolved spectroscopic properties of different families of galaxies
can be compared in an homogenous way. In particular, in this figure one sees 
(i) red and dry galaxies, with little or no gas, most of them luminous early
types; (ii) bluer and more gas rich galaxies, with a wider variety of
morphologies, and with extended starforming regions and (iii) galaxies clearly
dominated by AGN activity, with gas ionization concentrated in the central
regions ([NII]/H$\alpha$$>$1).

\section{Summary}

We have presented the CALIFA survey, the largest IFU survey currently being 
implemented. The results of this survey will allow us to make significant progress in 
many areas of galaxy evolution where large, single-fiber surveys are limited 
through a lack of spatial resolution and aperture biases. 
We will progressively report on the development of the survey in subsequent
articles, and through its webpage {\sc http://www.caha.es/CALIFA}.

\section*{Acknowledgments}   
%

We thank the {\it Viabilidad , Diseño , Acceso y Mejora } funding program ,
ICTS-2009-10 , and the {\it Plan Nacional de Investigaci\'on y Desarrollo}
funding program, AYA2010-22111-C03-03, of the Spanish {\it Ministerio de
  Ciencia e Innovacion}, for the support given to this project.
%

%

\begin{thebibliography}{}
\small
%
\bibitem{2009ApJS..182..543A} Abazajian, K.~N., et 
al.\ 2009, ApJS, 182, 543 


\bibitem{dezeeuw02}
{de Zeeuw}, P.~T., {Bureau}, M., {Emsellem}, E., {et~al.} 2002, MNRAS, 329,
  513

\bibitem{Kelz:2006p3341}
{Kelz}, A., {Verheijen}, M.~A.~W., {Roth}, M.~M., {et~al.} 2006,PASP, 118,
  129

\bibitem{RosalesOrtega:2010p3553} {Rosales-Ortega}, F.~F., {Kennicutt}, R.~C., {S{\'a}nchez}, S.~F., {et~al.}
  2010, MNRAS, 461

\bibitem{roth05}
{Roth}, M.~M., {Kelz}, A., {Fechner}, T., {et~al.} 2005, PASP, 117, 620

\bibitem{Sanchez:2004p2632}
{S{\'a}nchez}, S.~F. 2004, Astronomische Nachrichten, 325, 167

\bibitem{Sanchez:2006p331}
{S{\'a}nchez}, S.~F. 2006, Astronomische Nachrichten, 327, 850

\bibitem{sanchez07}
{S{\'a}nchez}, S.~F., {Cardiel}, N., {Verheijen}, M.~A.~W., {Pedraz}, S., \&
  {Covone}, G. 2007, MNRAS, 376, 125

\bibitem{sanchez10} S{\'a}nchez, S.~F., 
Rosales-Ortega, F.~F., Kennicutt, R.~C., Johnson, B.~D., Diaz, A.~I., 
Pasquali, A., \& Hao, C.~N.\ 2010, MNRAS, 1474 
%
%
\end{thebibliography}
\end{document}